\documentclass[twocolumn,tightenlines,aps]{revtex4}
\usepackage{graphicx}
\begin{document}
\title{Guiding molecules with electrostatic forces in Surface Enhanced Raman Spectroscopy (SERS)}

\author{P. D. Lacharmoise$\,^{1*}$ \protect\footnotetext{$*$ Electronic
address: placharmoise@icmab.es}}
\author{E. C. Le Ru$^2$}
\author{P. G. Etchegoin$^2$}

\affiliation{$^1$Institut de Ci\`encia de Materials de Barcelona-CSIC,\\Esfera UAB, 08193
Bellaterra, Spain}
\affiliation{$^2$The MacDiarmid Institute for Advanced Materials and Nanotechnology,\\ School of Chemical and Physical Sciences, Victoria University of Wellington,\\ P.O. Box 600 Wellington, New Zealand}

\begin{abstract}
In Surface Enhanced Raman Scattering (SERS) the problem of drawing molecules to the places where surface plasmon resonance enhancements will produce signals is one of the most basic ones, and the initial obstacle to every application of the effect. We explore the possibility of using electrostatic forces as a means to ``guide'' charged molecules in solution toward SERS active substrates. We also show explicitly the possibility of selectively enhancing different types of dyes according to their charge, and we discuss briefly possible extensions for other applications where ``electrostatic guiding'' could be an option.
\end{abstract}
\maketitle

\section{Introduction and overview}

The effect of applying a voltage to molecules adsorbed on metallic electrodes has been widely studied, essentially in the context of {\it electrochemistry}. By changing the potential of the working electrode with respect to a certain reference, the adsorbed molecules may change their oxidation state through redox reactions and/or modify their interaction with the surface. This topic is, in fact, at the core of what defines many electrochemical techniques and has an extensive and wide-ranging history that includes (among other things) dye molecules on metallic electrodes; two of the most common cornerstones of Surface Enhanced Raman Scattering (SERS) \cite{book}. Many standard dyes like nile blue (NB) \cite{PeLo90,NiFe90}, toluidine blue \cite{PeLo90,ScKa94,ScKa95}, meldola blue \cite{PeLo90,GoTo84,NaKa95}, brilliant cresyl blue \cite{PeLo90} and various derivatives of these parent structures have been studied under this approach, and successfully employed in chemically modified electrodes. Accordingly, in-situ SERS \cite{ChFu82,FlHi85,Pe91,Pe92,WeKi93} has allowed to follow these effects with high sensitivity down to the level of sub-monolayers of adsorbed species. SERS was used to obtain valuable information on the oxidation state of the molecules in the presence of an electric potential, of the orientation of the molecule when adsorbed on the surface, and of the electron transfer dynamics at the interface \cite{NiFe90,Pett82}. As a matter of fact, the discovery of SERS itself can be traced back to the field of electrochemistry \cite{1974FleischmannCPL,1977JeanmaireJEAC,1977AlbrechtJACS}; the symbiosis between the two fields has been widespread and fruitful.\\

The electrochemical problem in the presence of redox reactions is complex \cite{FlHi85}, and has more than one aspect that can be studied in detail. For example, the combination of electrochemistry, SERS, and biologically relevant molecules (like proteins) has led in the past to the study of electric field effects that include \cite{HilCSR}: $(i)$ protein dynamics, $(ii)$ redox-linked structural changes of conformations in proteins, $(iii)$ modulation of redox potentials, $(iv)$ reorganization energies, etc\ldots Moreover, a different approach to the use of electric potentials in SERS was introduced by Tian et al. \cite{TiLi96}, still within the boundaries of what is conventionally classified as electrochemistry. In order to obtain SERS spectra of surface species at electrode potentials where SERS active sites are normally unstable, they rapidly modulate the working electrode between two different potentials by applying a square-wave modulation voltage. The potential-averaged SERS spectrum contains the information on the surface species at the two modulated potentials. This form of electrochemical modulation \cite{TiLi96} has opened many new potential areas of study; not all of them yet fully explored.\\

All electrochemical studies put the emphasis on what actually happens once the molecule is on the electrodes. The potential is seen as a means to change the oxidation/reduction state of the molecule but, as a rule of thumb, never as a means to actually attract or selectively detect a specific molecule drawn from the solution. Electrodes are typically saturated with molecules through a series of volt-amperometric cycles that defines the initial state. This is followed subsequently by the study of the oxidation/reduction of the analytes as a function of an applied potential (with respect to a reference electrode) and in the presence of an underlying current through the cell (to drive the redox reaction).\\

Our motivation here is slightly different. Electric fields can be used in the first place to ``guide'' molecules toward the SERS active substrate. At a very basic level, one of the main (and first) practical limitations of SERS is how to draw together the {\it enhancement} with the {\it probe}. After all, it is the combination of the two that will produce a SERS signal. A wide variety of cases can be found in practice; from molecules that will show a natural affinity toward the substrate (through special chemical groups, for example) to molecules that will not stick to the metal at all due to, for example, unfavorable electrostatic interactions. A typical situation for water-soluble molecules is that of charged species dissolved in solution, whose attachment will strongly depend on the substrate charge. The possibility of using electric fields to selectively attract dyes of different charges seems therefore appealing. Here the interest is not in the electrochemical phenomena that happens at the electrodes but rather in the effects of the ``guiding field'' to increase (or select) the molecules to be monitored by SERS on a given substrate.\\

Several strategies have been either followed or proposed in the literature to solve the ``convolution problem'' of the dye concentration with the enhancement factor distribution, and these go from chemical functionalization of molecules (to attach them to the SERS substrate in the right places) to optical forces, that might affect the position and residence time of molecules at hot-spots. All these methods have advantages and disadvantages, and some of them (like optical forces) have not been yet conclusively demonstrated to be of practical use. Arguably, a rather more obvious choice in solution is the use of {\it electrostatic forces} to draw charged species toward target regions where they can profit from SERS enhancements. Hence, to explore this possibility we performed absorption/desorption cycles of different standard dyes for SERS applications. By attracting the molecules to the SERS active surface, we can measure the spectra of dyes that would normally be mostly non-absorbed. Experiments with two dyes of different charges in solution were also achieved, showing selective SERS spectra of each specimen depending on the direction of the applied field.
To characterize the effect of the electric field on the spatial concentration of the analytes in the cell, we studied first the distribution of the dyes by {\it fluorescence} experiments. We shall show that (under adequate experimental conditions) electrostatic forces may be implemented as an alternative way to selectively drive molecules into specific sites with high SERS activity, which is one of the main issues that concerns the the use of SERS at a very basic level.\\

\begin{figure}[ht]
  \begin{center}
   \includegraphics[angle=90,width=8cm]{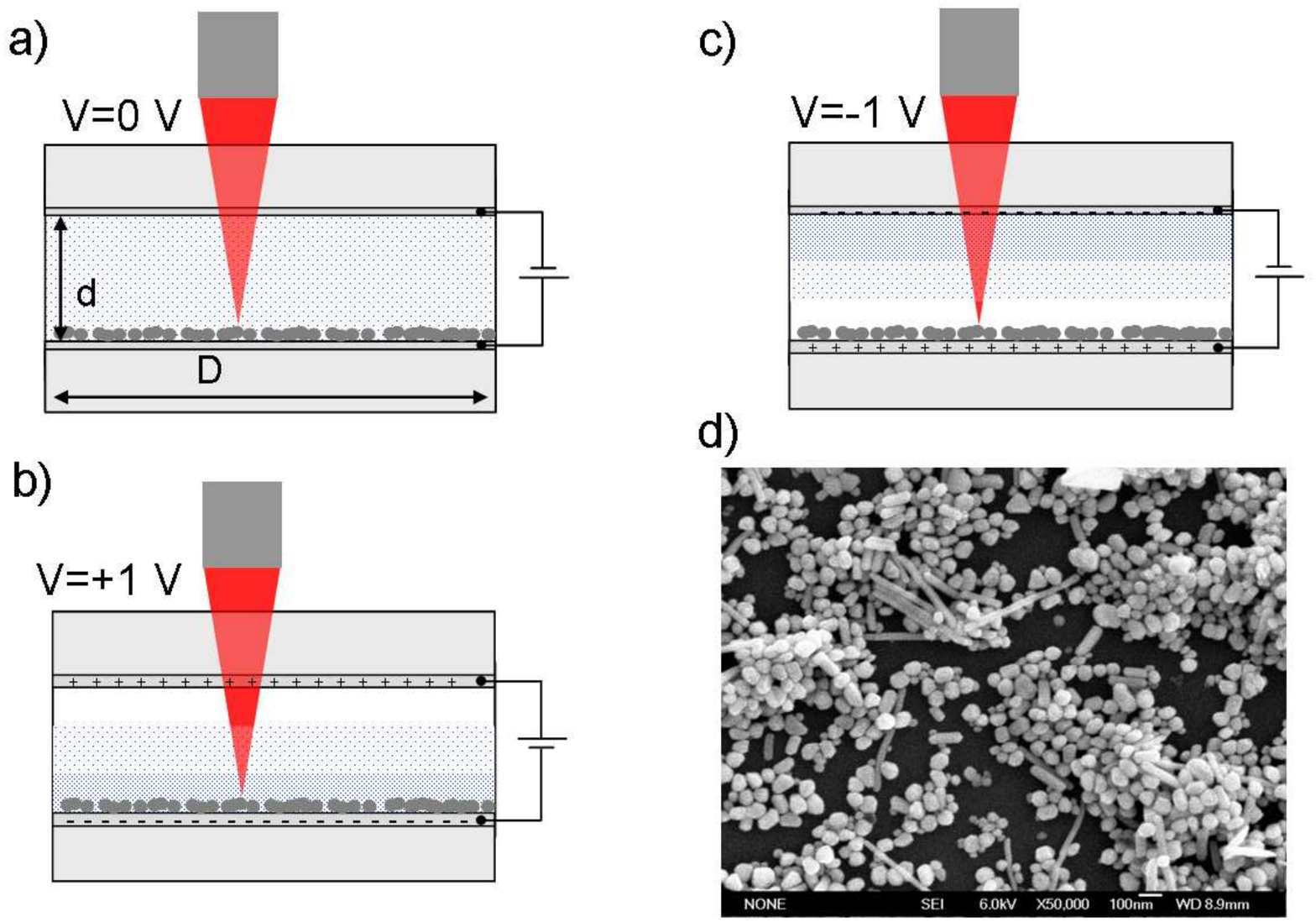}
   \caption{Electrolytic cell with height $d$ and diameter $D$. The distribution of analyte molecules for $V=0\,{\rm V}$ {\bf (a)}, $V=1\,{\rm V}$ {\bf (b)} and $V=-1\,{\rm V}$ {\bf (c)} is sketched for the case of a positively charged analyte (like nile blue, NB). In both fluorescence and SERS experiments the laser comes from the top of the cell and is focused either on a defined focal plane (parallel to the cell plates) or on the SERS substrate, respectively. {\bf (d)} SEM image of the sodium citrate reduced (Lee \& Meisel \cite{LeMe82}) Ag colloids that were used in SERS experiments. The colloids have an average diameter of $\sim 60\,{\rm nm}$ and form a dense mat on the surface of the bottom plate of the cell.}
\label{fig1}
\end{center}
\end{figure}

The paper is organized as follows: first, we describe the electrolytic cell and the SERS substrates used in our experiments. We then characterize the effect of the applied electric field and evidence the redistribution of charged molecules inside the cell by optical experiments. Finally, we present and discuss the SERS results for different situations, stressing the concept of selective attraction of charged species towards SERS active substrates.

\section{Experimental}

A simple electrolytic-cell (no reference electrode) in which it is possible to do optical measurements in the presence of an applied electrostatic field is used. A schematic representation of the cell is shown in Fig. \ref{fig1}. Two glass slides covered with a $200\,{\rm nm}$ film of ITO-glass are used as the capacitor plates. The separation $d$ of the plates is determined by a teflon spacer that surrounds a circular cell of diameter $D$, with the whole arrangement being press-fitted by external clamps. The studied cell had $d=1\,{\rm mm}$ and $D=6\,{\rm mm}$. A DC voltage source is connected to the ITO contacts. Being transparent in the visible range, the ITO layer allows us to illuminate the cell from the top with a $\times 50$ long working distance microscope objective. An applied voltage between the plates polarizes the solution giving rise to an electrostatic electric field perpendicular to the plates. The dyes are diluted in the cell in deionized ultra-pure water, but there is an inevitable residual amount of ions and counterions in the cell (starting from the analytes themselves). We shall not dwell into an analysis of the double-layer formation or composition, but rather only concentrate on the effect of the field on the spatial distribution of the analyte concentration (which we can monitor directly with a fluorescence scan in the axial direction). The expected profile of the analyte concentration when applying positive and negative voltages to the cell is schematically represented (for a positively charged dye in solution) in Fig. \ref{fig1}.\\

The SERS substrate on one of the surfaces of the cell was produced by previously drying a drop of a solution containing metallic colloids. The colloid layer was supported on the bottom cell plate (see Fig. \ref{fig1}), and remains stable (through its interaction with the ITO substrate) upon refilling the cell with water + analytes. Both, commercial gold colloids ($\sim$10nm in diameter, Sigma-Aldrich) and  freshly prepared silver colloids were used in our experiments. The latter were produced by sodium-citrate reduction following the standard Lee-\&-Meisel recipe \cite{LeMe82}. Colloids are typically concentrated by centrifugation and removal of the supernatant before drying them on the ITO glass, thus increasing their surface density and creating a fairly thick and homogeneous layer with high SERS activity. A SEM image of the obtained Ag colloids is shown in Fig. \ref{fig1}(d). We can observe a dense mat of Ag nanoparticles with an average overall size of $\sim 60\,{\rm nm}$ and multiple shapes. All the measurements were taken with the $\lambda$=633\,nm line of a He-Ne laser with a power density of $\sim$0.36\,mW/$\mu$m$^{2}$. As mentioned before, the beam is focused inside the cell through a $\times 50$ long working distance microscope objective with a numerical aperture (NA) of 0.5. The collected light is measured with a Jobin Yvon LabRam HR800 spectrometer, equipped with a N$_2$-cooled CCD. A grating with 300 lines/mm was chosen in order to acquire a sufficiently wide spectral window to contain the complete Raman spectrum of the studied molecules (which is superimposed, typically, on top of their fluorescence backgrounds). A Topward-6303D DC power supply was used as voltage source to polarize the cell.

\begin{figure}[ht]
  \begin{center}
   \includegraphics[angle=90,width=8cm]{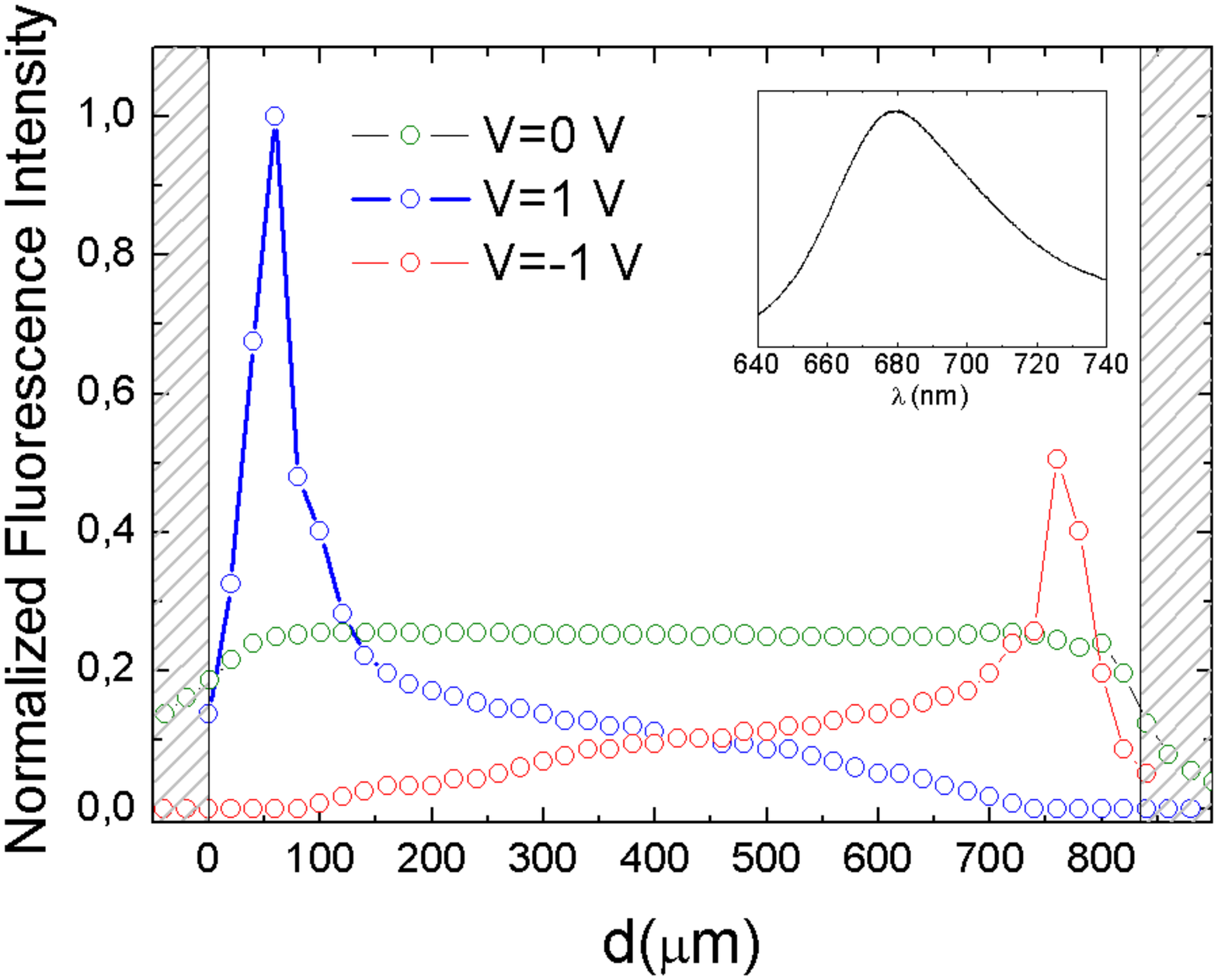}
   \caption{NB fluorescence profiles of a 100\,nM solution for $V=0$, 1 and $-1\,{\rm V}$. The profiles are obtained by an axial scan of the fluorescence signal with $20\,\mu{\rm m}$ steps. (inset) Fluorescence spectrum of NB excited at 633\,nm.}
\label{fig2}
\end{center}
\end{figure}

\section{Spatial distribution of dyes under electric fields}

In order to study how the electric field affects the distribution of charged molecules inside the cell, we scanned initially the fluorescence of a 100\,nM nile blue (NB) solution from the bottom plate to the top one at different voltages in a cell {\it without} colloids on the surface. NB,  a {\it positively charged} dye at ph$\sim$ 7, is in resonance at 633\,nm laser excitation and has an intense fluorescence spectrum centered at $\lambda=677\,{\rm nm}$ (see the inset of Fig. \ref{fig2}). Since the solution is homogeneous and no SERS substrates are used in these experiments, the intensity of the fluorescence is directly proportional to the quantity of probed molecules in the scattering volume, which is $\sim 100\,\mu{\rm m}^3$ within our experimental conditions. The scattering volume was experimentally determined by a combination of an axial intensity and confocal pinhole scans following the procedure described in full details in Ref. \cite{EFpaper}. We performed scans by moving the focal plane of the microscope in steps of $20\,\mu{\rm m}$ and acquiring a fluorescence spectrum at each point.  The density profiles obtained for different voltages are shown in Fig. \ref{fig2}. As expected, when $V=0$, the NB molecules are homogeneously distributed resulting in a constant fluorescence intensity across the cell. The fluorescence intensity decays slightly when approaching the electrodes because of the fact that the probed volume of solution diminishes as the scattering volume crosses the solution/electrode interface. When $V=1\,{\rm V}$ the NB distribution is strongly modified, as expected. It is clear that the NB cations leave the positively charged surface and are attracted toward the negative surface. The NB concentration profile can be separated into two regions depending on the distance $d$ from the negative electrode. From $d\sim 700$ (i.e. close to the positive electrode) to $\sim 140\,\mu{\rm m}$ the NB concentration increases approximately linearly as we approach the negative electrode. From $d\sim 140$ to $\sim 60\, \mu{\rm m}$ the density of molecules also grows, but at a much faster pace. This change of behavior can be associated with the beginning of the double layer region close to the boundary. It is important to note that the scans were performed at a sufficiently long time after applying the voltage, in order to reach a stationary state. This transient time (and the final steady-state density profile) depends on the magnitude of the applied voltage and the nature of the molecule, as we shall discuss later in the experiments with {\it two} different dyes. The fluorescence intensity decreases sharply close to the surface and this has, we believe, at least two different origins. Firstly, as it occurs with the density profile for the $V=0$ case, the probed volume decreases when the scattering volume crosses the interface. Moreover, the fluorescence of those molecules that are directly adsorbed to the surface may be quenched by non-radiative processes (or redistribution of the emission) involving the interaction between the molecule and the surface (ITO in this case). By integrating and comparing the areas underneath the density curves for $V=0$ and $V=1\,{\rm V}$ it is possible to estimate roughly the number of molecules adsorbed on the surface when the voltage is applied. The ratio of the areas for $V=0$ and $V=1\,{\rm V}$ indicates that $\sim 40\%$ of the NB molecules present in solution at $V=0$ are adsorbed to the surface after the voltage is applied ($V=1\,{\rm V}$). Note that the low NB concentration of 100\,nM used in these experiments ensures a sub-monolayer surface coverage (even for 100\% adsorption). When we switch the voltage to $V=-1\,{\rm V}$ the profile (after a characteristic transient time) is symmetrically inverted, as can be appreciated also in Fig. \ref{fig2}. The steady-state concentration distributions of the dyes in the volume of the cell is a balance between the random diffusion of the molecules and the drift imposed by the electric field (with a characteristic mobility that depends on the molecule).

\section{SERS experiments}

\begin{figure}[ht]
  \begin{center}
   \includegraphics[angle=0,width=8cm]{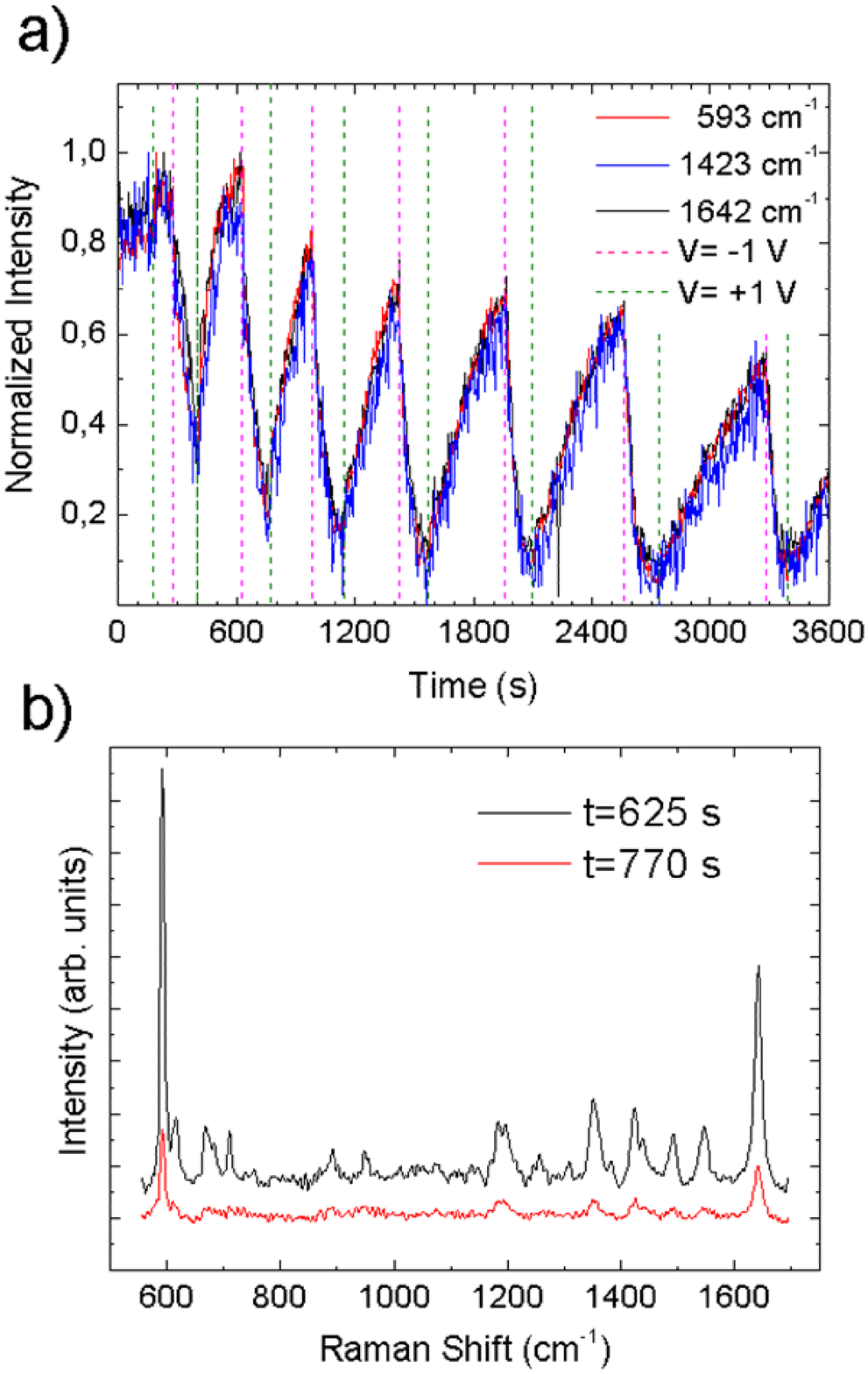}
   \caption{{\bf (a)} {\it In-situ} SERS at a fixed point on the bottom surface of the cell (Fig. \ref{fig1}(a)) of a $10\,\mu{\rm M}$ solution of NB during voltage cycles. The intensities of three NB Raman modes are displayed as a function of time (and normalized to their respective intensities at $t=0$). Commercial gold colloids dried on the bottom electrode were used as SERS substrate. {\bf (b)} Raman spectra taken from (a) at $t=625$ and $770\,{\rm sec}$. The fluorescence background was subtracted from each spectra to ease the direct comparison of the Raman peaks and their relative intensities for both cases.}
\label{fig3}
\end{center}
\end{figure}

As a second step, we now focus on the SERS experiments performed in the cell (i.e. with colloids deposited on one of the electrodes). The basic idea is to have a controlled way to drive the probed molecules to and from the SERS active substrate by applying a voltage. As stressed before, our emphasis is not on the electrochemical aspects of the signals at the electrodes, but rather on the ability to selectively attract/repulse and (more importantly) {\it replace} molecules with different charges at the electrodes. For as long as the potentials used do not exceed an irreversibility threshold (which can happen for certain adsorbed species \cite{Pett82}), our interest here is to evaluate the feasibility of directing the molecules where it matters the most for SERS.\\

The fluorescence experiments show that (as expected) the distribution of the molecules can be strongly affected by the presence of an electric field resulting from the applied voltage. We can therefore profit from this to selectively adsorb or desorb charged molecules of different charges from the SERS substrates. Firstly, by alternating the cell voltage between negative and positive potentials we can make absorption/desorption cycles and probe the dynamics of the experiment by {\it in-situ} SERS monitoring. Fig. \ref{fig3}(a) shows the SERS intensity of three different modes of NB at $10\,\mu{\rm M}$ as a function of time. Commercial gold colloids are used here as SERS substrate on the bottom electrode. The laser is focused on the substrate surface and we measure Raman spectra every 5\,sec. The SERS activity is fairly homogeneous in different points of the sample and the intensity of the Raman peaks is, to a very good approximation, proportional to the number of molecules adsorbed on the surface. Several cycles were performed with no significant loss of molecules in the process, except for a small continuous decrease in the maximum intensity from one cycle to the other attributed to photobleaching of the molecules adsorbed at the sites with highest electromagnetic enhancements \cite{EtRu06,RuEt06}. The signal can, however, be recovered to its original values if the spot is changed to a different position after a couple of cycles. Furthermore, as we can see in Fig. \ref{fig3}(a), the time evolution of different Raman modes of the molecule is identical. An important observation is that the {\it relative Raman intensities} of different peaks remain unchanged after several cycles. This suggests that no new species of NB are created in the electrode by redox reactions at this potential \cite{Pett82} and also that the adsorption of the molecules in the presence of the electric field is not preferentially oriented. The latter are two characteristic situations that can occur in electrochemistry and, in fact, they do occur at this potential level for other molecules in our cell (like rhodamine 6G, for example). Accordingly, the experimental conditions we are using do not produce any substantial electrochemical change in the normal SERS spectra of the analyte; our cell is still of the ``dielectric type'' with a negligible steady-state leakage current that does not drive any significant (or at least detectable through SERS) oxidation/reduction of the molecules. If desired, electrochemical effects related to oxidation/reduction of species at electrodes could be further reduced by an improved design of the electrolytic cell.
In Fig. \ref{fig3}(b) we show two spectra of NB taken from the second cycle in Fig. \ref{fig3}(a) when the intensity is maximum (at $t=625\,{\rm sec}$) and minimum (at $t=770\,{\rm sec}$). The fluorescence background was subtracted from the measured spectra to make the comparison of the Raman peaks more direct. More than $\sim 80\,$\% of the molecules are desorbed from the substrate at $t=770\,{\rm sec}$. Nevertheless, as we already discussed, the Raman spectrum remains unchanged except for the overall intensity.\\

\begin{figure}[ht]
  \begin{center}
   \includegraphics[angle=0,width=8cm]{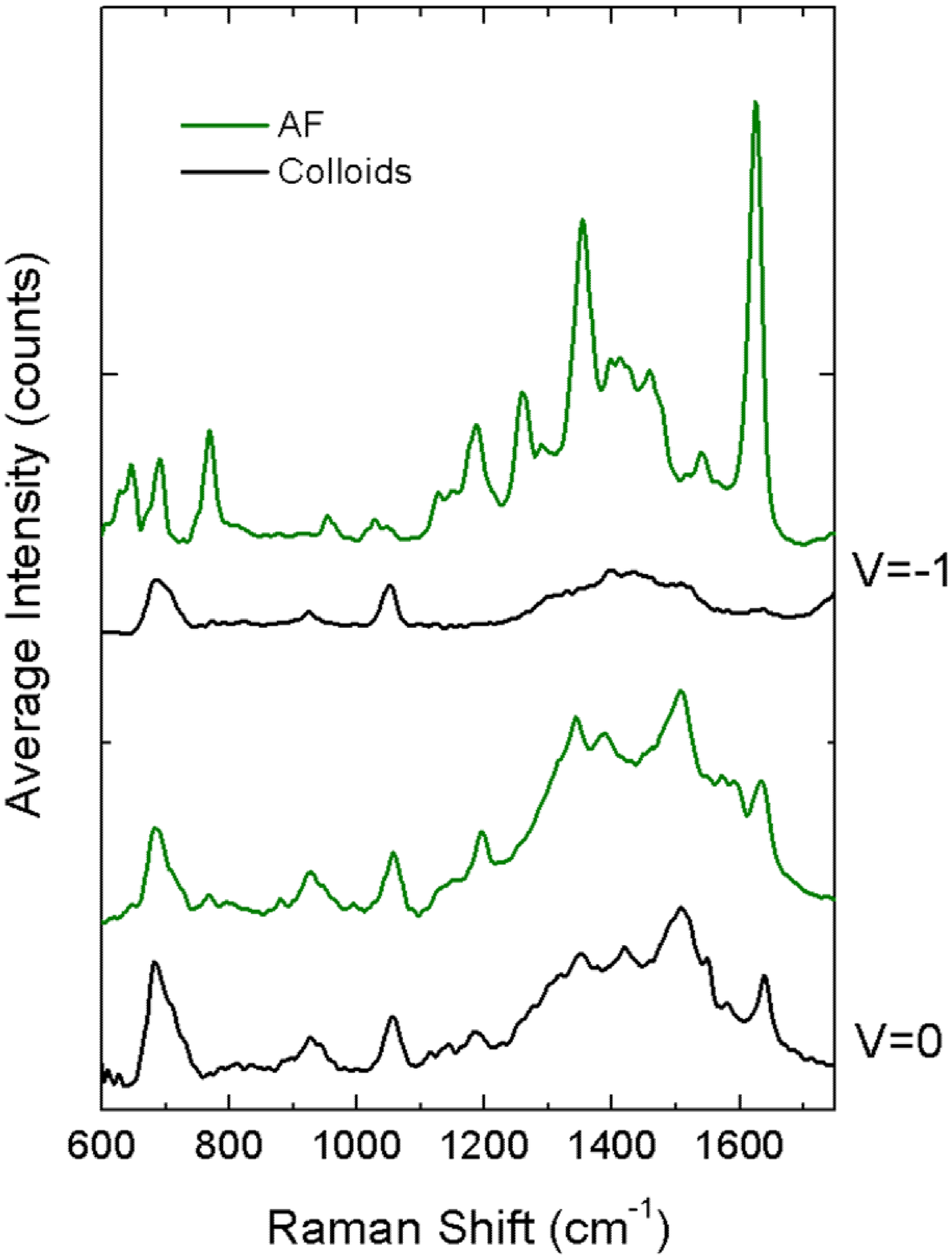}
   \caption{SERS spectra of a solution containing AF at $50\,\mu{\rm M}$ compared to the case of pure water for $V=0$ and $V=-1\,{\rm V}$. The experimental conditions were exactly the same in both cases. Note that the spectra coming purely from citrate (black curves), which is covalently bound to the colloids, does show a slight electrochemical difference in the relative intensities of peaks at $V=0$ and $-1\,{\rm V}$. This does not happen to the relative intensities of the peaks of the dye at any of the voltages we use. AF would {\it not} bind by itself to the SERS active substrate due to Coulomb repulsions between the negatively charged colloids and the the dye, which is also negatively charged in solution (by being a sulfonated dye). The SERS spectrum of the dye can only be observed if AF is electrostatically attracted to the colloids by the appropriate field. See the text for further details.}
\label{fig4}
\end{center}
\end{figure}

Moreover, it is also possible with this approach to drive molecules to substrates where they would not adsorb under normal conditions. Gold and silver colloids in solution are, in general, negatively charged \cite{colloids} and it is common to profit from the Coulomb attraction between positively charged molecules and the colloids to drive analyte adsorption. Conversely, SERS spectra of negative ions cannot be measured under normal conditions because of the poor surface adsorption due to Coulomb repulsion forces. However, it is possible to overcome this in the electrolytic-cell by applying the appropriate field. In Fig. \ref{fig4}, for example, the SERS spectra of a $50\,\mu{\rm M}$ solution of Alexa Fluor 488 (AF) (which is a well-known dye in biological applications, due to its high fluorescence quantum yield and high photo-stability). This molecule is sulfonated and exists in its anionic form in solution at ph$\sim 7$. Ag colloids like the ones shown in Fig. \ref{fig1}(d) were used in this case. Figure \ref{fig4} displays the spectra taken at $V=0$ and $V=-1\,{\rm V}$ after the transient time. In order to distinguish the signal of AF from other contributions (mainly due to the citrate molecules on the colloids), we compare each AF SERS spectrum with a SERS spectrum taken in exactly the same colloids, at the same voltages and times, but with {\it pure water} instead of the solution containing AF. Since both spectra are almost identical at $V=0$, it is clear that AF does not adsorb on the substrate (as expected) due to Coulomb repulsion. At $V=-1\,{\rm V}$, however, the spectra clearly show the Raman peaks attributed to AF. This is a, in fact, an important demonstration of the effect we want to highlight here; i.e. the possibility of selective adsorption of molecules on the SERS substrate using an electric field, not as an agent of oxidation/reduction, but rather as a guiding force to adsorb the dye where it matters. In the particular case of negatively charged dyes the usefulness is perhaps more evident, for we are able to observe a SERS signal from a substrate that was originally not suited at all to anion adsorption.

\begin{figure}[ht]
  \begin{center}
   \includegraphics[angle=0,width=8cm]{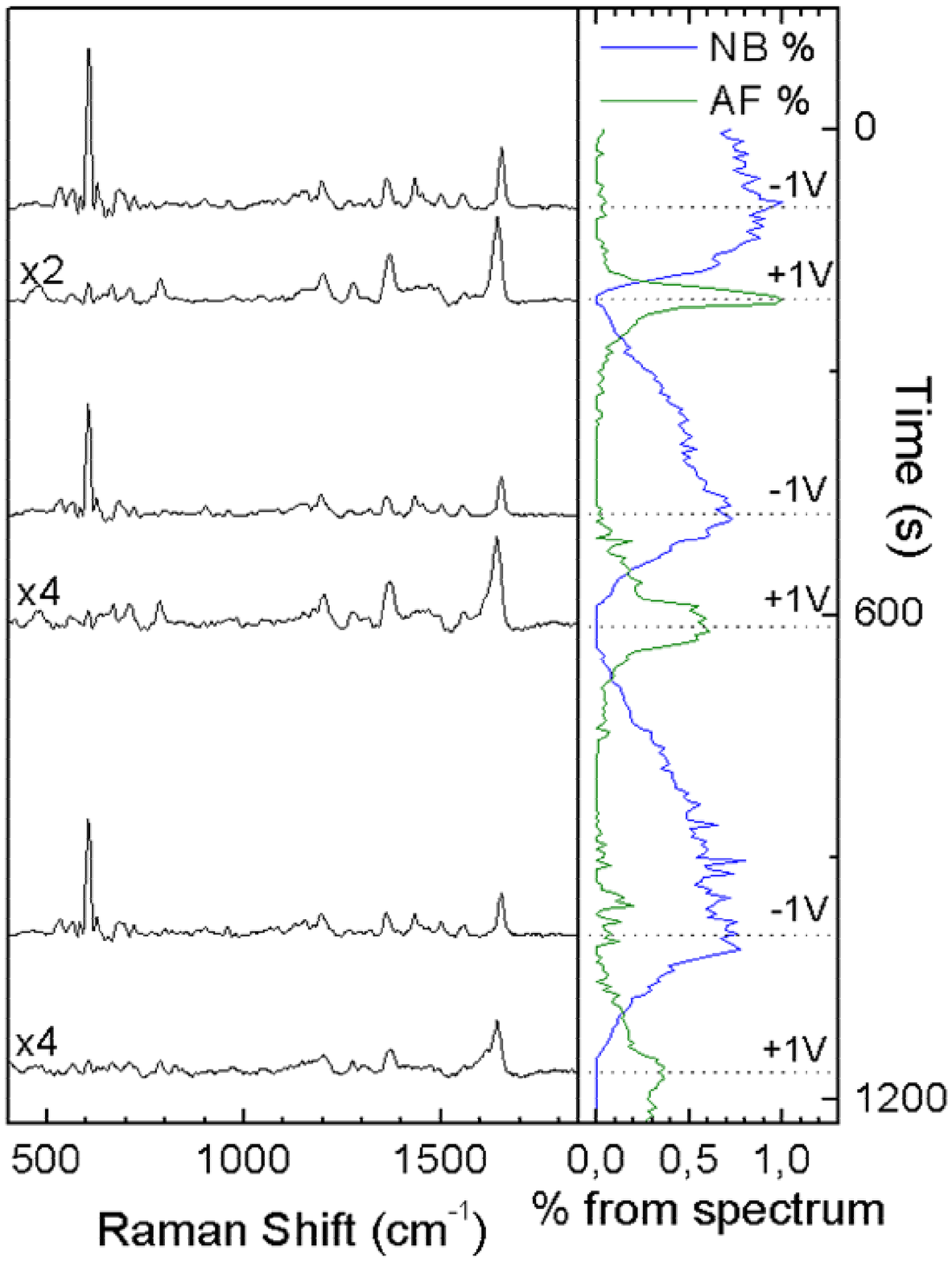}
   \caption{Two-analyte electrostatic binding experiment: the SERS spectrum of a solution containing NB at $500\,{\rm nM}$ and AF at $25\,\mu{\rm M}$ as cations and anions, respectively, is studied as a function of time. By alternating the voltage between the plates it is possible to selectively drive the charged ions to the active SERS substrate. The right panel displays the relative contributions of each dye to the total spectrum. The left side panel shows the Raman spectra obtained when each of the dyes are totally adsorbed to the SERS substrate (with no visible trace of the other one). The different ``response times'' of the two dyes to stick/unstick from the surface come from the combination of different surface interactions and intrinsic diffusions.}
\label{fig5}
\end{center}
\end{figure}

Along these lines, a very illustrative experiment consists in measuring a solution of molecules with opposite ionic charges that are selectively attracted to the active SERS substrate by applying a voltage with the appropriate polarity. We studied, for example, a solution with NB at $500\,{\rm nM}$ and AF at $25\,\mu{\rm M}$ as cations and anions, respectively. The concentration of each dye is chosen in order to have a comparable SERS signal in our experimental conditions (the molecules have very different intrinsic Raman cross sections at 633nm). We show in Fig. \ref{fig5} the results obtained for such a two-analyte experiment. As for the single dye experiments, we focus the laser beam on the SERS substrate (Ag Lee \& Meisel colloids in this case) and collect Raman spectra every $5\,{\rm sec}$. On the right-hand side in Fig. \ref{fig5} we can follow the evolution of the SERS intensity of each dye. The individual contributions are deconvoluted from the mixed spectra and the characteristic peaks of each species are integrated to obtain the relative quantities of adsorbed dyes during the cycles. As expected, only the NB molecules are adsorbed to the colloids at $t=0$ ($V=0$), because of the charge affinity. Note that this is despite the fact that the concentration of AF is $\sim$50 times larger than NB. At $t=95\,{\rm sec}$ we set the voltage to $-1\,{\rm V}$ and the NB molecules start to desorb from the SERS substrate. At $t=170\,{\rm sec}$ the AF contribution increases sharply; the adsorbed NB molecules are completely replaced by AF dyes at $t=210\,{\rm sec}$. Following this period, we then switch the voltage to $+1\,{\rm V}$ and the inverse process takes place: AF leaves the negatively charged surface and NB is readsorbed. The complete cycle is repeated three times in Fig. \ref{fig5}, to convey a general impression on how this selective attaching to the SERS substrate by electrostatic forces works. On the left-hand side of Fig. \ref{fig5} we can see the SERS spectra measured just before the voltage is switched. As can be appreciated, each spectra contain the fingerprint of the dye that was adsorbed during the last voltage period, confirming that the substrate active sites are completely occupied by the selected dye (no residual peaks of the other). Since NB is much more resonant that AF at $\lambda$=633\,nm, the latter has lower intrinsic SERS cross section; AF spectra were multiplied by a factor in the figure for clarity. Note also that the response timescales of the signal are clearly different for each molecule. This is attributed to a mixture of the specific interaction (affinity) of the molecule with the surface and their intrinsic mobility in the liquid.

\section{Conclusions}

The experimental results presented here, and in particular Fig. \ref{fig5}, demonstrate that --for as long as the electrochemical aspects of the problem are kept under control and oxidation/reduction of species is not a major issue-- electric fields can be used as an additional degree of freedom to attach molecules on SERS substrates and even selectively choose analytes from the solution.

There are several situations in SERS where an ``electrostatic guiding force'' for attracting the molecules to the SERS substrate could be useful or desirable. Arguably, it constitutes a much more direct ``handle'' on the analytes than other options that have been proposed in the literature (like optical forces \cite{2002XuPRL,2006SvedbergNL}), but this is undoubtedly at the expense of introducing additional potential complications into the problem (like possible oxidation/reduction of the species at the substrate).
The ultimate application of this principle would be to solve one of the outstanding issue of single-molecule
SERS \cite{PCCPfeature}, namely: positioning precisely the molecule at a hot-spot where it can be detected.
To this end, one could envision a situation where a single hot-spot at a differential potential with respect to a distant (insulated) electrode in solution acts as an ``electrostatic trap'' for guiding molecules toward it.
The hot-spot itself can be at the same potential, or can have two sections at a small potential difference floated with respect to the distant electrode. This could allow a combination of the ideas presented in this paper and in Ref. \cite{2008WardNL}. In that sense, we regard the results of this paper as a {\it demonstration of principle} of a degree of freedom that has not been hitherto widely used, but that can provide an additional tool to solve one of the most basic (and initial) problems in any SERS experiment; i.e. the convolution of the spatial distribution of the enhancement factor and the position of the analytes.


\acknowledgments PDL acknowledges an I3P-CSIC grant and the hospitality of the Victoria University of Wellington and the MacDiarmid Institute for Advanced Materials and Nanotechnology in New Zealand. ECLR and  PGE are indebted to the Royal Society of New Zealand (RSNZ) for financial support through a {\it Marsden grant}. Special thanks are give to Chris Galloway (Victoria University) for providing his background removal program (COBRA) for some of the data analysis.

\end{document}